# Numerical simulation of α-quartz under non-hydrostatic compression. Memory glass and five-coordinated crystalline phases

James Badro[*], Jean-Louis Barrat[†]and Philippe Gillet[‡]

## Abstract

The behavior of α-quartz under hydrostatic and non-hydrostatic high-pressure conditions has been investigated in Molecular Dynamics (MD) simulations of silica in order to clarify the role of non-hydrostatic stresses in the amorphization process. It is shown that the amorphization threshold is not modified if the stress along the c direction is lowered, so that the mean amorphization pressure can effectively be lowered under non-hydrostatic conditions. On the other hand, the application of a positive uniaxial stress along the c axis, results in the appearance of a new crystalline phase, where all silicon atoms are in five-fold coordination.

[*]Laboratoire de Sciences de la Terre (URA 726), Ecole Normale Supérieure de Lyon, 46 Allée d'Italie F-69364 Lyon Cedex 07, FRANCE

[†]Département de Physique des Matériaux (URA 172), Université Claude Bernard, 43 bd. du 11 novembre 1918, F-69622 Villeurbanne Cedex, FRANCE

[‡]Laboratoire de Sciences de la Terre (URA 726), Institut Universitaire de France, Ecole Normale Supérieure de Lyon, 46 Allée d'Italie, F-69364 Lyon Cedex 07, FRANCE

Previous experimental and theoretical studies have shown that α-quartz, the stable form of silica under standard conditions, undergoes a crystalline to amorphous transition at high pressure and room temperature [1, 2, 3, 4, 5, 6, 7, 8, 9, 10]. It has since been proposed that the application of pressure together with the inhibition of thermal activation led to a mechanical instability [6, 8] in the crystalline network driving the crystalline to amorphous transition. It was proposed that non-hydrostatic stresses played an important role in the amorphization process [4, 11] and it was shown in a recent experimental study on the high-pressure behavior of α-$AlPO_4$ [12] that these stresses are of a fundamental importance in the crystal to amorphous phase transition of this material.

In this letter, molecular dynamics (MD) simulations of the simple silica model recently developped in reference [13] are used to explore the effect of non-hydrostaticity on the crystal to amorphous transformation. This interatomic potential [13], obtained by *ab-initio* calculations on $H_4SiO_4$ clusters, is chosen for its ability to reproduce a number of structural properties of α-quartz, among which pressure induced amorphization at 22 GPa and 300 K [7]. The simulations are carried out in a modified (N,P,T) isobaric-isothermal canonical ensemble, in which the stresses $\sigma_{xx}, \sigma_{yy}, \sigma_{zz}$ can be varied independently. The simulation box is rectangular, with the z direction parallel to the **c** axis of α quartz and contains 1440 atoms (480 $SiO_2$ units). The external stress is monitored using Berendsen's scaling [14] independently in the three directions. The equations of motion are integrated with a time step of 1.34 fs. The densities of states reported in this paper are calculated from the (mass weighted) power spectrum of the velocity autocorrelation functions, computed in the (N,V,E) microcanonical ensemble. They are smoothed by a gaussian distribution of width 20 $cm^{-1}$, in order to have the same resolution as those obtained by inelastic neutron scattering experiements.

A reference amorphous state was first created by high pressure amorphization under hydrostatic conditions. Amorphization occured at 22 GPa, as previously reported [7, 8, 10]. That amorphous system was then compressed to 25 and 30 GPa. All three systems were allowed to age for 14 ps at constant P and T and for another 14 ps at constant V and T after which the vibrational densities of states (DOS) reported in figure 1 were computed. These systems were then decompressed to 0 GPa during 130 ps and were aged during 14 ps at constant V and T once equilibrium was attained. The DOS were once again calculated and are also reported in figure 1. Coordination distribution before the transition (up to 21 GPa) was identical to that of α-quartz, with all silicon atoms being four-coordinated with respect to oxygen. Table 1 lists the average Si-O and O-Si coordination numbers above 22 GPa. The higher the pressure above the transition, the greater the proportion of six-coordinated silicons and three-coordinated oxygens. It can also be noticed that four bonded silicons and two bonded oxygens constitute the large majority of the decompressed amorphous material. This information is of prime interest because it is not possible to analyze $SiO_2$ *in-situ* by X-ray absorption spectroscopy due to the low energy of the silicon K-edge and high absorption of the diamonds in this energy domain. It is also clear that the higher the final pressure, the more the local environment upon decompression differs from the original material as far as coordination is concerned. The densities of state clearly reveal the growth of bands around 1000 $cm^{-1}$ and between 600 and 700 $cm^{-1}$ when pressure is raised after amorphization. The latter phenomenon has already been observed by raman scattering on silica glass upon pressurization [15] and by infrared absorption performed both on α-quartz and silica glass [3]. Upon decompression, the spectra reproduce qualitatively the infrared measures previously reported by Williams *et al.* [3], at least in the high frequency range where the growth of the 900 $cm^{-1}$ line can be observed. One should be very careful when analyzing the low frequency domain, since the finite size of the simulation box can affect the vibrational spectrum. All together, these results confirm the remarkable ability of the model for describing silica under high pressure.

Two different compression paths were then chosen for the application of non-hydrostatic stresses. First, the stress parallel to the **c** axis ($\sigma_{zz}$) was kept constant and below 22 GPa and the stresses in the perpendicular plane ($\sigma_{xx}$ and $\sigma_{yy}$) were raised up to amorphization. Then, $\sigma_{xx}$ and $\sigma_{yy}$ were maintained at a constant value below 22 GPa and $\sigma_{zz}$ was increased.

In the first case, amorphization always occured for $\sigma_{xx}$ and $\sigma_{yy}$ between 22 and 23 GPa, hence the needed stress in the (**a**,**b**) plane for driving the phase transition does not seem to depend on the value of $\sigma_{zz}$, at least in the range where the calculations were done. These runs were made with several values of $\sigma_{zz}$, namely 21, 20, 19, 18, 15 and 13 GPa. A close study of the radial distribution functions reveals that the very non-hydrostatic runs with low values of $\sigma_{zz}$ contained a small part of crystalline material. Si-O and O-Si coordination numbers were calculated in all six cases and are reported in table 1 for the most non-hydrostatic run. Upon decompression, all systems except the ones obtained under the most non-hydrostatic conditions remained amorphous, and completely similar to those obtained under hydrostatic conditions. The latter systems ($\sigma_{zz}$=15 and 13 GPa) underwent a partially reversible transition in that sense that the recovered sample consisted of back-transformed quartz with some defects. The structure thus obtained is qualitatively similar to the one reported in figure 4. The Si-O and O-Si coordination numbers as well as the vibrational density of state were calculated and compared with those of $\alpha$-quartz (table 1 and figure 2). The first interpretation of these results for $\alpha$-quartz, which is more compressible in the **a** and **b** directions than following the **c** axis, is that amorphization seems to happen as the result of an instability occuring in this plane, or at least in a plane which has a small dihedral angle with the latter [10, 11]. It also seems clear the the mean pressure of amorphization (*i.e.* the mean value of the three stresses) is lower than the pressure needed to hydrostatically amorphize the system. This is the first direct observation of this phenomenon, although most experimental studies seem to agree on this point. Amorphization is partially reversible upon decompression if the average non-hydrostatic pressure is low enough, and it could be that the crystalline domains at high pressure act as favorable nucleation sites upon decompression, driving the amorphous-to-quartz transition.

In the second series of runs, $\sigma_{xx}$ and $\sigma_{yy}$ were kept constant and $\sigma_{zz}$ was raised until amorphization. This time, no direct quartz to amorphous phase transition was observed. The system underwent a crystal-to-crystal transition towards a new phase, in which all silicon atoms are five-coordinated (see table 1). A slice parallel to the $\alpha$-quartz **c** axis is represented in figure 3. The pressure conditions at which this structure appeared were $\sigma_{xx} = \sigma_{yy}$ =20 GPa and $\sigma_{zz}$ =27 GPa. Pressure was isotropically raised from 0 to 20 GPa, followed by an increase to 27 GPa along the **c** axis. Each compression was carried out over 50000 steps accounting for a total compression time of 134 ps.

In order to test the stability of this structure, its temperature was raised at constant pressure, without observable variation up to 1500 K. Upon further compression parallel to the **c** axis ($\sigma_{zz}$ =29 GPa), the sample became partially amorphous but in this case only five and six-coordinated silicon atoms were present in the disordered network (see table 1).

The decompression phase was the exact opposite of the compression phase; the excess non-hydrostatic stress was first released and only then was the pressure lowered isotropically from 20 GPa. In the case of the new crystalline phase, a phase transition occured at 2 GPa and the structure reverted back to the original quartz structure, whereas the partially amorphous system transformed once again quasi-reversibly to quartz containing defect sites at 10 GPa (fig. 4). The corresponding DOS is reported in figure 2 and the coordination distribution in the table.

This study clearly illustrates the importance of non-hydrostaticity in the solid state amorphization phenomenon. From a practical viewpoint, the most important conclusion is that the mean amorphization pressure can be considerably lowered in the presence of non-hydrostatic stresses. The amorphous solids that are produced under moderately non-hydrostatic conditions are very similar to those obtained in an hydrostatic environment. The situation becomes qualitatively different under strongly non-hydrostatic conditions (i.e. when the stress along the **c** axis differs by more than 30% of the other stresses). For small $\sigma_{zz}$, amorphization is imperfect and the amorphous state reverts back to crystal upon decompression ("memory glass"). It is interesting to note that the recovered material's vibrational density of state (fig. 2) presents broader spectral lines and weeker signal due to the apparition of defects upon recrystallization. These two observations have been reported experimentally in the case of $\alpha$-$AlPO_4$ berlinite [12] and anorthite [16] which both recrystallize upon decompression, and could therefore mean that these recovered crystalline phases contain some defects. For high $\sigma_{zz}$, amorphization is pre-emted by a new crystal to crystal phase transition. The high pressure phase can amorphize, but again the resulting amorphous state is a "memory glass" that transforms back to quartz upon decompression. It has already been shown in a hydrostatic experiment [5] that phase transitions to a metastable crystalline form could constitute a pathway to solid state amorphization. This new phase, obtained under non-hydrostatic conditions could be another example of such crystal-to-crystal phase transitions.

These results are very suggestive of the great potential richness of solid state amorphization under non-hydrostatic conditions. It should be emphazised, however, that they were obtained using a very simple model for silica, and very high compression rates. In particular, the existence of a new crystalline phase (which would be the first stable five-coordinated polymorph of silica), and of reversible crystal to amorphous transitions in silica, should be regarded as interesting possibilities, but are not established by such simulations. Carefully designed non-hydrostatic anvil cell experiments will be necessary to investigate such effects.

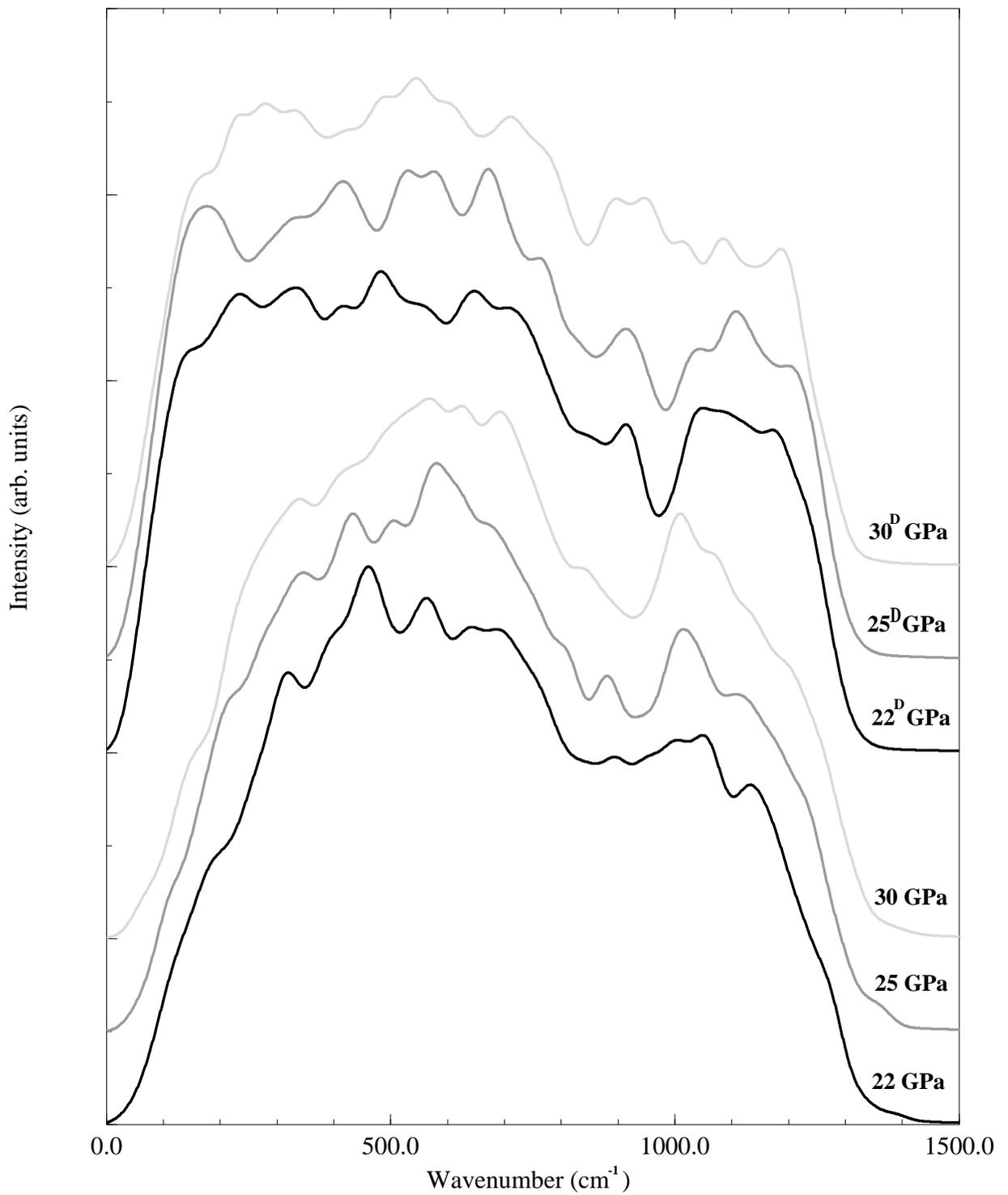

Figure 1: The vibrational density of state upon compression and decompression in a hydrostatic environment. The pressures are reported on the figure . The spectra with the D superscript correspond to systems decompressed from the reported pressure.

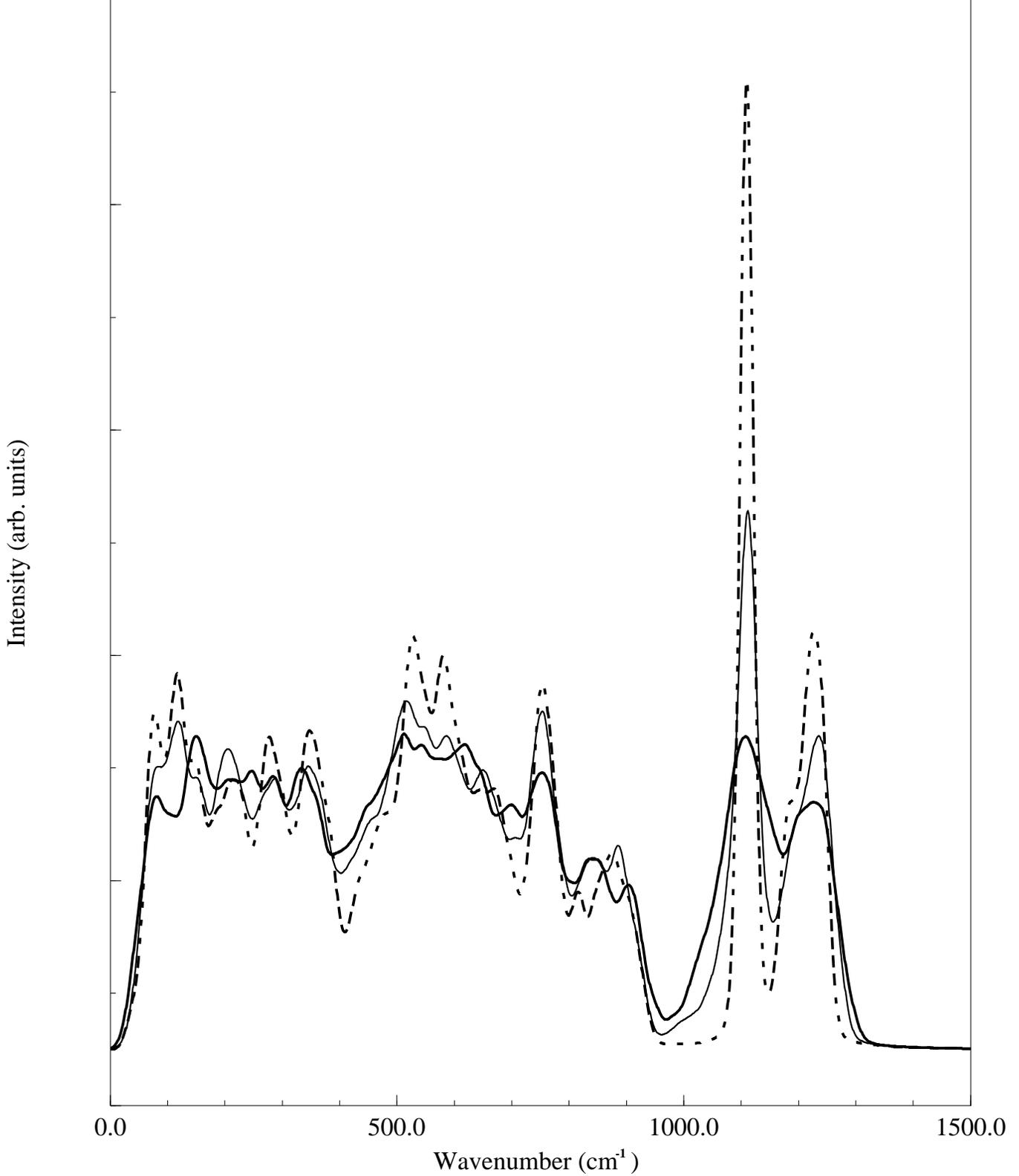

Figure 2: The vibrational density of state of a system decompressed from $\sigma_{xx}=\sigma_{yy}=22$ GPa and $\sigma_{zz}=13$ GPa (thick solid line) and from $\sigma_{xx}=\sigma_{yy}=20$ GPa and $\sigma_{zz}=29$ GPa (thin solid line) compared with that of $\alpha$-quartz at ambiant conditions (dotted line). The decompression in both cases is partly reversible. The sample back-transforms into crystalline quartz but contains some defects. Coordination of all three phases are almost the same (see table 1). This figure shows that the effect of defects present in the crystalline lattice is far from being negligible as far as the intensity and the width of the spectral lines are concerned.

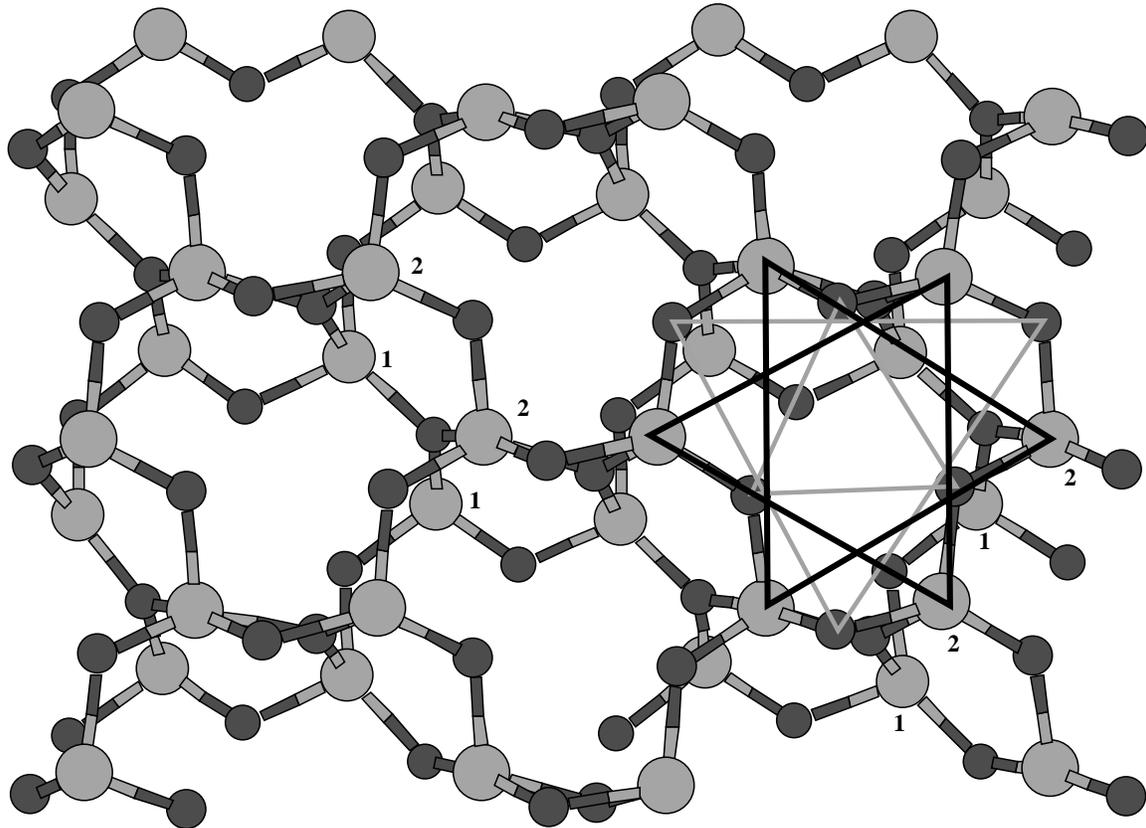

Figure 3: The new high-pressure phase viewed down the **c** axis. The silicon atoms (large light grey spheres) positioned on the edges of the two similar equilateral triangles (4.5 angstroms of side length) are in the same plane and the distance between these triangles is apporximately 1 angstrom in the **c** direction. The oxygen atoms (small dark grey spheres) are disposed on the edges of two non-identical equilateral triangles. The atoms on the edges of these triangles constitue a sheet representing one third of the cell in the **c** direction. The two other sheets are superposed as to form a helix and the helical step is worth approximately 3 angstroms. Two sheets can be seen on this slice (indicated by the numbers near the silicon atoms). The equilateral triangle formed by the projection of the three helically superposed silicon atoms on the (**a**,**b**) plane has a side length of 1.1 angstroms.

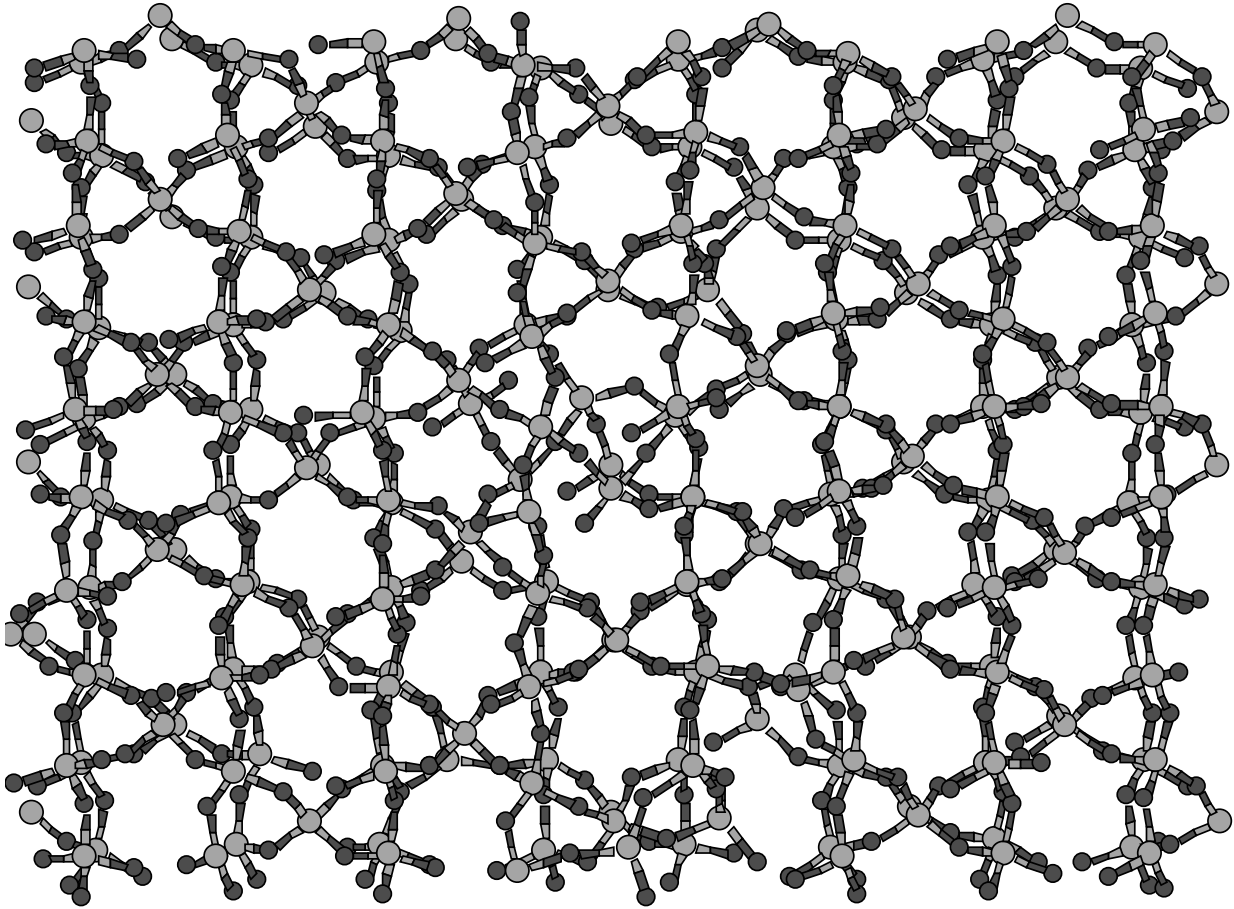

Figure 4: A slice cut perpendicular to the **c** axis of the system decompressed from (20,20,29) GPa. The decompressed sample recrystallizes as quartz but contains some defects. The silicon atoms are represented as large light grey spheres whereas the oxygen atoms are the small dark grey spheres. This is a typical example of a "memory" glass.

Table 1: Si-O and O-Si coordination distribution in hydrostatic and non-hydrostatic runs. The pressures followed by letter D correspond to a system at ambient pressure decompressed from the indicated pressure. In the first column, pressures represented with three figures in parentheses represent the $\sigma_{xx}$, $\sigma_{yy}$ and $\sigma_{zz}$ components of the stress, respectively. The last column lists the system's phase in the corresponding pressure conditions.

| P (GPa) | $Si^{IV}$ | $Si^V$ | $Si^{VI}$ | $O^{II}$ | $O^{III}$ | Phase |
|---|---|---|---|---|---|---|
| 0 | 100% | 0% | 0% | 100% | 0% | $\alpha$-quartz |
| 22 | 12.5% | 42% | 45.5% | 34.5% | 65.5% | amorphous |
| 25 | 8% | 38.5% | 53.5% | 30% | 70% | amorphous |
| 30 | 7% | 36.5% | 56.5% | 28% | 72% | amorphous |
| 22 D | 75% | 23.5% | 1.5% | 87% | 13% | amorphous |
| 25 D | 69% | 26% | 5% | 82% | 18% | amorphous |
| 30 D | 56% | 36% | 8% | 73% | 27% | amorphous |
| (22,22,13) | 13% | 34% | 53% | 32% | 68% | amorphous |
| (22,22,13) D | 97% | 3% | 0% | 98% | 2% | $\alpha$-quartz + defects |
| (20,20,27) | 0% | 100% | 0% | 50% | 50% | new phase |
| (20,20,27) D | 100% | 0% | 0% | 100% | 0% | $\alpha$-quartz |
| (20,20,29) | 0% | 90% | 10% | 45% | 55% | new phase + amorphous |
| (20,20,29) D | 99% | 1% | 0% | 99% | 1% | $\alpha$-quartz + defects |